\documentclass[dvips]{plain}
\usepackage{supertabular,lscape,epsfig}
\usepackage{amssymb}
\usepackage{amsmath}
\usepackage{wasysym}

\SetPages{1}{0}

\SetVol{0}{2022}

\begin{document}

\begin{Titlepage}
\Title{Search for planets in hot Jupiter systems with multi-sector TESS photometry. II. Constraints on planetary  companions in 12 systems.}
\Author{G.~~M~a~c~i~e~j~e~w~s~k~i}
{Institute of Astronomy, Faculty of Physics, Astronomy and Informatics,
         Nicolaus Copernicus University in Toru\'n, Grudziadzka 5, 87-100 Toru\'n, Poland,
         e-mail: gmac@umk.pl}

\Received{January 2022}
\end{Titlepage}

\Abstract{Uninterrupted observations from space-borne telescopes provide the photometric precision that is required to detect shallow transits of small planets missed by ground-based surveys. We used data from the Transiting Exoplanet Survey Satellite (TESS) to search for nearby planetary companions in 12 planetary systems with hot Jupiters: HD~2685, Qatar-10, WASP-4, WASP-48, WASP-58, WASP-91, WASP-120, WASP-121, WASP-122, WASP-140, XO-6, and XO-7. We also applied the transit timing method based on homogeneously determined mid-transit times in order to search for non-transiting companions that could gravitationally perturb the already known planets. We found no additional planets in those systems down to the regime of sub-Neptunian globes. This negative result is in line with statistical studies, supporting the high-eccentricity migration as a pathway of the investigated giant planets to the tight orbits observed today.}{Hot Jupiters -- Stars: individual: HD~2685, Qatar-10, WASP-4, WASP-48, WASP-58, WASP-91, WASP-120, WASP-121, WASP-122, WASP-140, XO-6, XO-7 -- Planets and satellites: individual: HD~2685~b, Qatar-10~b, WASP-4~b, WASP-48~b, WASP-58~b, WASP-91~b, WASP-120~b, WASP-121~b, WASP-122~b, WASP-140~b, XO-6~b, XO-7~b}


\section{Introduction}

Although hot Jupiters constitute an arbitrarily defined group of massive planets on tight orbits, they are clearly in contrast to giant planets on wider orbits if the occurrence rate in compact planetary systems is considered. The vast majority of hot Jupiters are found to be lonely, while about 50\% of warm Jupiters (with orbital periods above ten days) have nearby planetary companions (Huang \etal 2016). Recent statistics show that only $7.3^{+15.2}_{-7.3}$\% of hot-Jupiter systems might have a compact orbital architecture (Hord \etal 2021). This finding is meaningful for theories of planetary formation, favouring a high eccentricity migration (Rasio \& Ford 1996) as the main pathway of the hot Jupiters origin.

In Maciejewski (2020), we presented the motivation, methodology, and the first results of our project, the aim of which is to search for nearby planetary companions to hot Jupiters. We apply the transit detection method to the photometric time series acquired with the Transiting Exoplanet Survey Satellite (TESS, Ricker \etal 2014). The multi-sector data allow us to probe the sub-Neptune radius regime down to about 2 Earth radii ($R_{\oplus}$) for orbital periods up to 10 days. In addition, our study includes a homogenous timing analysis for transits of the known planets. This feature allows us to search for gravitational perturbations induced by non-transiting companions.

In this paper, we report on results obtained for 12 systems: HD 2685 (Jones \etal 2019), Qatar-10 (Alsubai \etal 2019), WASP-4 (Wilson \etal 2008), WASP-48 (Enoch \etal 2011), WASP-58 (H\'ebrard \etal 2013), WASP-91 (Anderson \etal 2017), WASP-120 (Turner \etal 2016), WASP-121 (Delrez \etal 2016), WASP-122 (Turner \etal 2016), WASP-140 (Hellier \etal 2016), XO-6 (Crouzet \etal 2017), and XO-7 (Crouzet \etal 2020).


\section{Systems of the sample}

The observational properties of the examined systems are summarised in Table~1. Their stellar and planetary properties are collected in Table~2. Below, short characterisation of the individual systems is provided.

\MakeTable{lcccccc}{12.5cm}{Observational properties of the systems of the sample.} 
{\hline
System  & RA (J2000)  & Dec (J2000) & $m_{\rm G}$  & Distance & $d_{\rm tr}$        & $\delta_{\rm tr}$ \\
        & hh:mm:ss.s  & $\pm$dd:mm:ss   & (mag)    & (pc)     & (min)                 & (ppth)            \\
\hline 
HD 2685  & 00:29:19.0 & --76:18:15 &  9.5 & $198.0\pm0.8$ & $268.2^{+3.8}_{-4.1}$ & $8.89^{+0.08}_{-0.07}$ \\
Qatar-10 & 18:57:46.6 &  +69:34:15 & 12.7 & $568.2\pm6.2$ & $164.3^{+3.8}_{-3.8}$ & $14.19^{+0.15}_{-0.16}$ \\
WASP-4   & 23:34:15.1 &  -42:03:41 & 12.3 & $269.2\pm3.8$ & $130.1^{+1.1}_{-1.4}$ & $22.7^{+0.3}_{-0.3}$ \\
WASP-48  & 19:24:39.0 &  +55:28:23 & 11.5 & $460.2\pm4.6$ & $186.2^{+7.3}_{-7.0}$ &  $8.69^{+0.12}_{-0.13}$ \\
WASP-58  & 18:18:48.3 &  +45:10:19 & 11.6 & $368.3\pm2.8$ & $221.0^{+5.8}_{-6.6}$ & $12.7^{+0.2}_{-0.2}$ \\
WASP-91  & 23:51:22.9 & --70:09:10 & 11.6 & $292.9\pm1.8$ & $141.8^{+3.2}_{-3.5}$ & $14.37^{+0.21}_{-0.18}$ \\
WASP-120 & 04:10:27.9 & --45:53:54 & 10.9 & $385.4\pm3.3$ & $210^{+11}_{-10}$     & $5.57^{+0.08}_{-0.08}$ \\
WASP-121 & 07:10:24.1 & --39:05:51 & 10.4 & $272.0\pm1.6$ & $171.3^{+1.6}_{-2.1}$ & $13.94^{+0.13}_{-0.11}$ \\
WASP-122 & 07:13:12.4 & --42:24:35 & 10.8 & $251.9\pm1.6$ & $130.5^{+2.3}_{-2.6}$ & $12.10^{+0.10}_{-0.10}$ \\
WASP-140 & 04:01:32.5 & --20:27:04 & 10.8 & $118.0\pm0.6$ &  $91.6^{+1.9}_{-1.9}$ & $18.61^{+0.32}_{-0.26}$ \\
XO-6     & 06:19:10.4 &  +73:49:40 & 10.4 & $237.1\pm2.5$ & $178.3^{+4.1}_{-3.6}$ & $11.55^{+0.08}_{-0.08}$ \\
XO-7     & 18:29:54.9 &  +85:14:00 & 10.5 & $235.7\pm1.2$ & $168.3^{+5.3}_{-5.0}$ &  $8.72^{+0.09}_{-0.09}$ \\
\hline
\multicolumn{7}{l}{Coordinates were taken from the Gaia Data Release 2 (DR2, Gaia Collaboration \etal 2018).}  \\
\multicolumn{7}{l}{$m_{\rm G}$ is the apparent brightness in the $G$ band from DR2. Distance is calculated on DR2}  \\
\multicolumn{7}{l}{parallaxes. $d_{\rm tr}$ and $\delta_{\rm tr}$ are the transit duration and transit depth, both refined in this study.}  \\
\multicolumn{7}{l}{ppth stands for parts per thousand of normalised out-of-transit flux.}  \\
}

\MakeTable{lccccccc}{12.5cm}{Physical properties of the systems of the sample.} 
{\hline
System  & $T_{\rm eff}$ (K) & $\log g_{\star}$ & [Fe/H] (dex) & $R_{\star}$ $(R_{\odot})$ & $M_{\rm b}$ $(M_{\rm Jup})$ & $R_{\rm b}$ $(R_{\rm Jup})$ & Source\\
\hline 
HD 2685  & $6801\pm76$ & $4.21\pm0.03$ & $+0.02\pm0.06$ & $1.56\pm0.05$ & $1.17\pm0.12$ & $1.44\pm0.05$ & (1)\\
Qatar-10  & $6124\pm46$ & $4.303\pm0.027$ & $+0.02\pm0.09$ & $1.254\pm0.026$ & $0.736 \pm 0.090$ & $1.543 \pm 0.040$ & (2)\\
WASP-4  & $5500\pm150$ & $4.45^{+0.02}_{-0.03}$ & $0.0\pm0.2$ & $0.94^{+0.04}_{-0.03}$ & $1.215^{+0.087}_{-0.079}$ & $1.416^{+0.068}_{-0.043}$ & (3)\\
WASP-48 & $5920\pm150$ & $4.03\pm0.04$ & $-0.12\pm0.12$ & $1.75 \pm 0.09$ & $0.98 \pm 0.09$ & $1.67 \pm 0.10$ & (4)\\
WASP-58 & $5800\pm150$ & $4.27\pm0.09$ & $-0.45\pm0.09$ & $1.17 \pm 0.13$ & $0.89 \pm 0.07$ & $1.37 \pm 0.20$ & (5)\\
WASP-91 & $4920\pm90$ & $4.49\pm0.03$ & $+0.19\pm0.13$ & $0.86 \pm 0.03$ & $1.34 \pm 0.08$ & $1.03 \pm 0.04$ & (6)\\
WASP-120 & $6450\pm120$ & $4.035\pm0.049$ & $-0.05\pm0.07$ & $1.87\pm0.11$ & $4.85 \pm 0.21$ & $1.473 \pm 0.096$ & (7)\\
WASP-121 & $6460\pm140$ & $4.2\pm0.2$ & $+0.13\pm0.09$ & $1.46 \pm 0.03$ & $1.183^{+0.064}_{-0.062}$ & $1.807 \pm 0.039$ & (8)\\
WASP-122 & $5720\pm130$ & $4.3\pm0.1$ & $+0.32\pm0.09$ & $1.52 \pm 0.03$ & $1.284 \pm 0.032$ & $1.743 \pm 0.047$ & (7)\\
WASP-140 & $5300\pm100$ & $4.51\pm0.04$ & $+0.12\pm0.10$ & $0.87 \pm 0.04$ & $2.44 \pm 0.07$ & $1.44^{+0.42}_{-0.18}$ & (9)\\
XO-6 & $6720\pm100$ & $4.04\pm0.10$ & $-0.07\pm0.10$ & $1.93 \pm 0.18$ & $1.9 \pm 0.5$ & $2.07 \pm 0.22$ & (10)\\
XO-7 & $6250\pm100$ & $4.246\pm0.023$ & $+0.432\pm0.057$ & $1.480 \pm 0.022$ & $0.709 \pm 0.034$ & $1.373 \pm 0.026$ & (11)\\
\hline
\multicolumn{8}{l}{$T_{\rm eff}$, $g_{\star}$, [Fe/H], and $R_{\star}$ are the effective temperature, gravitational acceleration in cgs, metallicity, and radius for a host }  \\
\multicolumn{8}{l}{star, respectively. $M_{\rm b}$ and $R_{\rm b}$ are the mass and radius of a transiting planet in Jupiter units. Data sources are: }  \\
\multicolumn{8}{l}{(1) -- Jones \etal (2019), (2) -- Alsubai \etal (2019), (3) -- Wilson \etal (2008), (4) -- Enoch \etal (2011), } \\
\multicolumn{8}{l}{(5) -- H\'ebrard \etal (2013), (6) -- Anderson \etal (2017), (7) -- Turner \etal (2016), (8) -- Delrez \etal (2016), }  \\
\multicolumn{8}{l}{(9) -- Hellier \etal (2016), (10) -- Crouzet \etal (2017), (11) -- Crouzet \etal (2020).}  \\
}

The planet HD~2685~b was a second hot giant planet detected in TESS observations (Jones \etal 2019). Its orbital period is 4.1 days. The Doppler data show that the orbit is circular at $2\sigma$ level. The planet is recognised as being inflated, likely due to the high irradiation received from its early F-type host star.

Transits of Qatar-10 b were detected by the Qatar Exoplanet Survey (Alsubai \etal 2019). The planet orbits its F/G host star on a circular orbit within 1.6 days. No follow-up studies of this system are available in the literature.

WASP-4~b was a fourth planet announced by the Wide Angle Search for Planets (WASP) consortium and the first one discovered by that project on the southern hemisphere (Wilson \etal 2008). It orbits a G7 dwarf within $1.3$ days. The system's parameters were refined in the follow-up studies by Gillon \etal (2009), Winn \etal (2009), Southworth \etal (2009), Sanchis-Ojeda \etal (2011), Dragomir \etal (2011), Nikolov \etal (2012), and Petrucci \etal (2013). The spectroscopic observations of the Rossiter-McLaughlin (RM) effect by Triaud \etal (2011) showed that the planet's orbit is aligned with the stellar rotation axis. This finding was confirmed and refined by modelling subtle signatures of starspot occultations in transit light curves (Sanchis-Ojeda \etal 2011). Hoyer \etal (2013) used the timing of those occultations to determine the star's rotation period, which was found to be equal to about 34 days. Using the NASA/ESA Hubble Space Telescope, Ranjan \etal (2014) found that the emission spectrum of the planet is consistent with a model of a carbon-rich atmosphere. Huitson \etal (2017) analysed the optical transmission spectrum acquired from the ground and found that the atmosphere of WASP-4~b is cloudy. This finding was confirmed with independent observations by May \etal (2018) and Bixel \etal (2019). Intriguing results of a transit timing analysis based on TESS observations were reported by Bouma \etal (2019). The transits were observed about 80 s earlier than a previous transit ephemeris predicted them. This orbital period shortening was confirmed by Southworth \etal (2019) and Baluev \etal (2020). Bouma \etal (2020) demonstrated that the observed change in the orbital period is likely caused by a line-of-sight acceleration of the system's barycentre due to the presence of a 10--300 $M_{\rm Jup}$ companion on a 10--100 au wide orbit. However, Turner \etal (2021) used a broader data set and showed that the system is not accelerating uniformly towards the Earth. Instead, a massive planetary companion on a seven au orbit was detected. The presence of that additional planet cannot explain the observed change in the orbital period of WASP-4~b, leaving the question about a mechanism behind it opened.

WASP-48 b orbits a slightly evolved F-type star within 2.1 days (Enoch \etal 2011). The system parameters were verified by Sada \etal (2012) and Ciceri \etal (2015). The timing of occultations shows that the orbit is circular (O'Rourke \etal 2014, Clark \etal 2018). Murgas \etal (2017) used the transit spectroscopy method to show that a transmission spectrum of the planet is relatively flat and without features from atmospheric species. 

WASP-58 b was found around a metal-poor analogue of the Sun (H\'ebrard \etal 2013). Its orbital period very close to 5 days makes follow-up observations challenging from the ground. Mallonn \etal (2019) used amateur transit light curves to refine a transit ephemeris. No other photometric time series are available in the literature.

With its orbital period of 2.8 days, WASP-91 b is a typical hot Jupiter orbiting a K3 dwarf (Anderson \etal 2017). No follow-up studies of this system are available in the literature.

WASP-120~b belongs to a small group of massive hot giant planets on non-circular orbits (Turner \etal 2016). The orbital eccentricity was found to be $e_{\rm b} = 0.057^{+.022}_{-0.018}$, \ie non-zero with a significance of 3.3$\sigma$. The orbital period is 3.6 days. The system parameters were refined by Alexoudi (2022).

Due to its short orbital period of 1.3 days, WASP-121~b was announced as a hot Jupiter close to tidal disruption (Delrez \etal 2016). That resulted in numerous follow-up studies. The orbit is significantly miss-aligned, being just 1.15 times larger than the Roche limit of the planet. In the atmosphere of the planet, the signatures of the H$_2$O molecule and neutral atoms of H, K, Li, Mg, Na, Ca, Cr, Fe, Ni, and V were found (Evans \etal 2016, Gibson \etal 2020, Hoeijmakers \etal 2020, Borsa \etal 2021). The atmosphere has a stratosphere, \ie an outer layer with a temperature inversion (Evans \etal 2017), and an exosphere extending beyond the Roche lobe size (Sing \etal 2019). Based on TESS data, an orbital phase light curve reveals that heat is inefficiently transported from the dayside to the planet's nightside (Bourrier \etal 2020, Daylan \etal 2021). 

A giant planet on a 1.7-day orbit around the G2--G4 star WASP-122 was discovered by the WASP consortium (Turner \etal 2016), and it was independently confirmed by the Kilodegree Extremely Little Telescope (as KELT-14, Rodriguez \etal 2016). Using TESS data from Sector 7, Wong \etal (2020) redetermined the system parameters and found no significant phase shift in the brightness modulation in an orbital phase light curve.

WASP-140 b is a hot Jupiter on a 2.2 day orbit with a non-zero eccentricity of $e_{\rm b} = 0.047 \pm 0.004$ (Hellier \etal 2016). It produces grazing transits. The only follow-up studies come from Alexoudi (2022), who refined the system parameters.

XO-6 b (Crouzet \etal 2017) orbits a fast-rotating F5 dwarf within 3.8 days. Its orbit was found to be slightly misaligned with a measured sky-projected obliquity of about $-21^{\circ}$. Garai \etal (2020) refined system parameters and detected a periodic variation in transit times. However, the claimed signal was not confirmed with TESS observations (Ridden-Harper \etal 2020).

The XO-7 system comprises a hot Jupiter on a 2.9-day orbit, a wide-orbit companion with a minimum mass of 4 $M_{\rm Jup}$, and a G0 main-sequence host star (Crouzet \etal 2020). No follow-up studies of this system are available in the literature.


\section{TESS observations and data reduction}

We followed the procedure described in detail in Maciejewski (2020) to produce photometric time series from raw TESS images. In brief words: The short-cadence (SC) data, \ie with the exposure time of 2 minutes, were downloaded from the exo.MAST portal\footnote{https://exo.mast.stsci.edu}. The long-cadence (LC) data, downloaded at 30 minute cadence, were cut from full-frame images with the TESSCut\footnote{https://mast.stsci.edu/tesscut/} tool (Brasseur \etal 2019). The Lightkurve v1.9 package (Lightkurve Collaboration 2018) was used to produce final fluxes, which were de-trended, visually inspected, and normalised to unity outside the transits. The photometric noise rate (pnr, Fulton \etal 2011) was calculated to quantify data quality for each target and each sector. A list of observations used for individual targets is given in Table~3. 

\MakeTable{cccccccc}{12.5cm}{Details on TESS observations used.} 
{\hline
Sect./ & from--to & $pnr$ & $N_{\rm tr}$ & Sect./ & from--to & $pnr$ & $N_{\rm tr}$ \\
/Mode  &          &       &              & /Mode  &          &       &              \\
\hline 
\multicolumn{4}{c}{HD 2685}                & \multicolumn{4}{c}{WASP-91} \\
 1/SC & 2018-Jul-25--2018-Aug-22 & 1.0 &  7 &  1/SC & 2018-Jul-25--2018-Aug-22 & 2.4 & 10 \\
27/SC & 2020-Jul-04--2020-Jul-30 & 1.2 &  6 & 27/SC & 2020-Jul-04--2020-Jul-30 & 2.4 &  8 \\
28/SC & 2020-Jul-30--2020-Aug-26 & 1.0 &  6 & 28/SC & 2020-Jul-30--2020-Aug-26 & 2.5 &  9 \\
\multicolumn{3}{r}{total:}             & 19 & \multicolumn{3}{r}{total:}             & 27 \\
\multicolumn{4}{c}{Qatar 10}                & \multicolumn{4}{c}{WASP-120} \\
14/LC & 2019-Jul-18--2019-Aug-15 & 4.9 & -- &  4/SC & 2018-Oct-18--2018-Nov-15 & 2.3 &  6 \\
15/LC & 2019-Aug-15--2019-Sep-11 & 5.0 & -- &  5/SC & 2018-Nov-15--2018-Dec-11 & 2.1 &  6 \\
16/LC & 2019-Sep-11--2019-Oct-07 & 5.3 & -- & 30/SC & 2020-Sep-22--2020-Oct-21 & 2.3 &  8 \\
17/LC & 2019-Oct-07--2019-Nov-02 & 5.6 & -- & 31/SC & 2020-Oct-21--2020-Nov-19 & 2.2 &  7 \\
18/LC & 2019-Nov-02--2019-Nov-27 & 5.2 & -- & \multicolumn{3}{r}{total:}             & 27 \\
19/LC & 2019-Nov-27--2019-Dec-24 & 5.0 & -- & \multicolumn{4}{c}{WASP-121} \\
20/LC & 2019-Dec-24--2020-Jan-21 & 5.3 & -- &  7/SC & 2019-Jan-07--2019-Feb-02 & 1.9 & 17 \\
21/LC & 2020-Jan-21--2020-Feb-18 & 4.8 & -- & 33/SC & 2020-Dec-17--2021-Jan-13 & 1.9 & 19 \\
22/LC & 2020-Feb-18--2020-Mar-18 & 4.8 & -- & 34/SC & 2021-Jan-13--2021-Feb-09 & 1.8 & 19 \\
23/LC & 2020-Mar-18--2020-Apr-16 & 5.6 & -- & \multicolumn{3}{r}{total:}             & 55 \\
24/SC & 2020-Apr-16--2020-May-13 & 4.9 & 15 & \multicolumn{4}{c}{WASP-122} \\
25/SC & 2020-May-13--2020-Jun-08 & 4.9 & 14 &  7/SC & 2019-Jan-07--2019-Feb-02 & 1.9 & 13 \\
26/SC & 2020-Jun-08--2020-Jul-04 & 4.7 & 14 & 33/SC & 2020-Dec-17--2021-Jan-13 & 1.7 & 14 \\
40/SC & 2021-Jun-24--2021-Jul-23 & 4.7 & 16 & 34/SC & 2021-Jan-13--2021-Feb-09 & 1.8 & 14 \\
41/SC & 2021-Jul-23--2021-Aug-20 & 4.5 & 15 & \multicolumn{3}{r}{total:}             & 41 \\
\multicolumn{3}{r}{total:}             & 74 & \multicolumn{4}{c}{WASP-140} \\
\multicolumn{4}{c}{WASP-4}                  &  4/SC & 2018-Oct-18--2018-Nov-15 & 1.8 &  8 \\
 2/SC & 2018-Aug-22--2018-Sep-20 & 3.6 & 18 &  5/SC & 2018-Nov-15--2018-Dec-11 & 1.5 & 10 \\
28/SC & 2020-Jul-30--2020-Aug-26 & 4.5 & 17 & 31/SC & 2020-Oct-21--2020-Nov-19 & 1.7 & 10 \\
29/SC & 2020-Aug-26--2020-Sep-22 & 4.3 & 18 & \multicolumn{3}{r}{total:}             & 28 \\
\multicolumn{3}{r}{total:}             & 53 & \multicolumn{4}{c}{XO-6}  \\
\multicolumn{4}{c}{WASP-48}                 & 19/SC & 2019-Nov-27--2019-Dec-24 & 1.3 &  6 \\
14/LC & 2019-Jul-18--2019-Aug-15 & 3.3 & -- & 20/SC & 2019-Dec-24--2020-Jan-21 & 1.4 &  6 \\
15/LC & 2019-Aug-15--2019-Sep-11 & 3.2 & -- & 26/SC & 2020-Jun-08--2020-Jul-04 & 1.6 &  6 \\
16/LC & 2019-Sep-11--2019-Oct-07 & 3.1 & -- & 40/SC & 2021-Jun-24--2021-Jul-23 & 1.3 &  6 \\
23/SC & 2020-Mar-18--2020-Apr-16 & 3.4 & 10 & \multicolumn{3}{r}{total:}             & 24 \\
26/SC & 2020-Jun-08--2020-Jul-04 & 3.1 & 12 & \multicolumn{4}{c}{XO-7}  \\
40/SC & 2021-Jun-24--2021-Jul-23 & 2.9 & 12 & 18/LC & 2019-Nov-02--2019-Nov-27 & 1.6 & -- \\
41/SC & 2021-Jul-23--2021-Aug-20 & 2.6 & 12 & 19/LC & 2019-Nov-27--2019-Dec-24 & 1.6 & -- \\
\multicolumn{3}{r}{total:}             & 46 & 20/LC & 2019-Dec-24--2020-Jan-21 & 1.6 & -- \\
\multicolumn{4}{c}{WASP-58}                 & 25/SC & 2020-May-13--2020-Jun-08 & 1.7 &  9 \\
14/LC & 2019-Jul-18--2019-Aug-15 & 3.3 & -- & 26/SC & 2020-Jun-08--2020-Jul-04 & 1.7 &  8 \\
25/SC & 2020-May-13--2020-Jun-08 & 2.9 &  5 & 40/SC & 2021-Jun-24--2021-Jul-23 & 1.5 & 10 \\
26/SC & 2020-Jun-08--2020-Jul-04 & 3.1 &  5 & \multicolumn{3}{r}{total:}             & 27 \\
40/SC & 2021-Jun-24--2021-Jul-23 & 3.0 &  6 &  &  &  &  \\
\multicolumn{3}{r}{total:}             & 16 &  &  &  & \\\hline
\multicolumn{8}{l}{Mode specifies long cadence (LC) or short cadence (SC) photometry.}  \\
\multicolumn{8}{l}{$pnr$ is the photometric noise rate in parts per thousand (ppth) of the normalised flux per minute of}     \\
\multicolumn{8}{l}{ exposure. $N_{\rm tr}$ is a number of transits qualified for this study.}     \\
}


\section{Data analysis and results}

\subsection{Transit modelling}

\begin{figure}[thb]
\begin{center}
\includegraphics[width=1.0\textwidth]{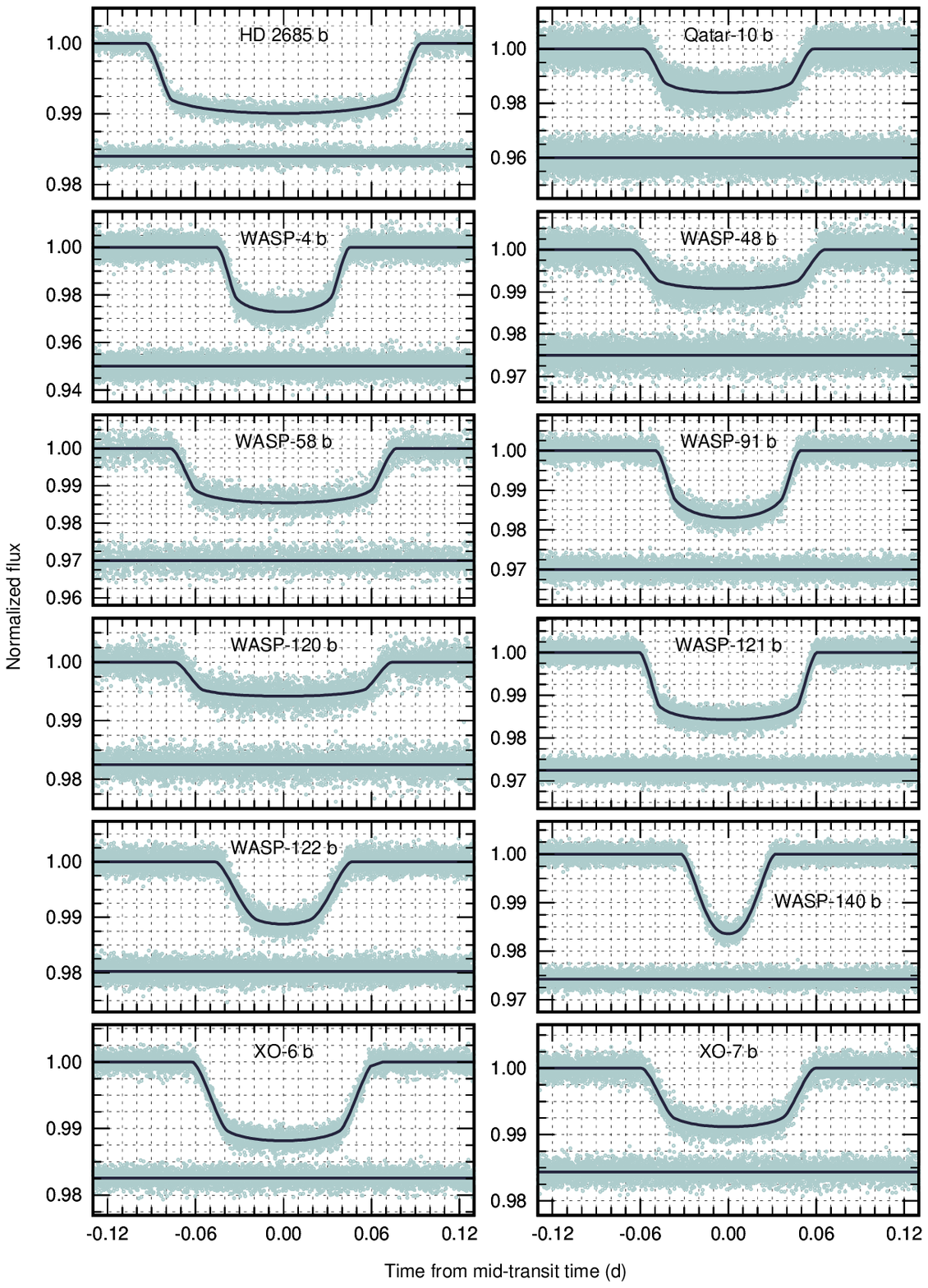}
\end{center}
\FigCap{Phase folded transit light curves with the best-fitting models for the planets of our sample. The residuals are plotted below each light curve.}
\end{figure}

The individual transits were extracted from the SC time series with margins of twice the transit duration around mid-transit times predicted by a trial ephemeris. Due to a lower time resolution, the LC time series provide diluted transits that affect a final transit model. Therefore they were skipped in further steps. The best-fitting transit models were obtained with the Transit Analysis Package (TAP, Gazak \etal 2012) in the configuration adopted from Maciejewski (2020). In trial iterations, the quadratic law's limb-darkening (LD) coefficients were free. Only for three systems: HD 2685, WASP-4, and WASP-121, the photometric time series were precise enough to determine at least the linear LC coefficients with uncertainties below 0.1. For the remaining systems, the uncertainties were higher but consistent with the theoretical predictions within $1\sigma$. Thus, the theoretical predictions from tables of Claret \& Bloemen (2011) were used and allowed to vary under the Gaussian penalty with a conservative value of 0.1. For planets on non-circular orbits, \ie WASP-120~b and WASP-140~b, the orbital eccentricity and the argument of periastron were allowed to vary under the Gaussian penalty taken from the literature. A mid-transit time was determined for each light curve, while the remaining transit model parameters were linked together. The best-fitting models for individual systems are plotted over the phase folded transit light curves in Fig.~1. The model parameters and the literature values are collected in Table~4.

\MakeTable{lccccc}{12.5cm}{System parameters from transit light curve modeling and from the literature.}
{\hline
Planet                     & $R_{\rm{p}}/R_{\star}$	         & $a/R_{\star}$			 & $i_{\rm{orb}}$ $(^{\circ})$ & $u_1$        & $u_2$                  \\
\hline
HD 2685 b & $0.09430^{+0.00042}_{-0.00038}$ & $7.56^{+0.10}_{-0.11}$    & $88.36^{+0.63}_{-0.50}$     & $0.29^{+0.05}_{-0.05}$    & $0.07^{+0.09}_{-0.10}$ \\
\multicolumn{1}{r}{Jones \etal (2019)} & $0.09467^{+0.00033}_{-0.00028}$ & $7.697^{+0.069}_{-0.054}$ & $89.25^{+0.42}_{-0.44}$     & $0.152^{+0.048}_{-0.042}$ & $0.40^{+0.15}_{-0.11}$ \\
Qatar 10 b                 & $0.11912^{+0.00065}_{-0.00068}$ & $4.780^{+0.095}_{-0.095}$ & $85.09^{+0.63}_{-0.58}$     & $0.27^{\rm a)}$           & $0.30^{\rm a)}$        \\
\multicolumn{1}{r}{Alsubai \etal (2019)} & $0.1265^{+0.0010}_{-0.0010}$    & $4.90^{+0.12}_{-0.12}$    & $85.87^{+0.96}_{-0.96}$     &                           &                        \\
WASP-4 b                   & $0.1506^{+0.0010}_{-0.0010}$    & $5.406^{+0.046}_{-0.056}$ & $88.94^{+0.71}_{-0.85}$     & $0.27^{+0.06}_{-0.07}$    & $0.45^{+0.17}_{-0.12}$ \\
\multicolumn{1}{r}{Wong \etal (2020)} & $0.1523^{+0.0010}_{-0.0010}$    & $5.438^{+0.044}_{-0.057}$ & $89.8^{+1.4}_{-1.4}$        & $0.374^{+0.065}_{-0.072}$ & $0.22^{+0.16}_{-0.14}$ \\
\multicolumn{1}{r}{Turner \etal (2021)} & $0.1516^{+0.0006}_{-0.0006}$    & $5.410^{+0.088}_{-0.088}$ & $88.02^{+0.69}_{-0.69}$     & $0.38^{\rm b)}$           & $0.21^{\rm b)}$        \\
WASP-48 b                  & $0.09324^{+0.00065}_{-0.00069}$ & $4.66^{+0.12}_{-0.11}$    & $82.03^{+0.51}_{-0.50}$     & $0.26^{\rm a)}$           & $0.30^{\rm a)}$        \\
\multicolumn{1}{r}{Enoch \etal (2011)} & $0.098^{+0.001}_{-0.001}$       & $4.23^{+0.24}_{-0.19}$    & $80.09^{+0.88}_{-0.79}$     &                           &                        \\
\multicolumn{1}{r}{Murgas \etal (2017)} & $0.09330^{+0.00088}_{-0.00088}$ & $4.76^{+0.08}_{-0.08}$    & $82.47^{+0.34}_{-0.34}$     &                           &                        \\
WASP-58 b                  & $0.11254^{+0.00088}_{-0.00078}$ & $11.28^{+ 0.28}_{-0.32}$  & $88.73^{+0.75}_{-0.53}$     & $0.28^{\rm a)}$           & $0.28^{\rm a)}$        \\
\multicolumn{1}{r}{H\'ebrard \etal (2013)} & $0.120^{+0.004}_{-0.004}$       & $10.3^{+1.2}_{-1.2}$      & $87.4^{+1.5}_{-1.5}$        &                           &                        \\
WASP-91 b                  & $0.11986^{+0.00086}_{-0.00082}$ & $9.43^{+0.18}_{-0.20}$    & $87.51^{+0.31}_{-0.32}$     & $0.45^{\rm a)}$           & $0.24^{\rm a)}$        \\
\multicolumn{1}{r}{Anderson \etal (2017)} & $0.1225^{+0.0012}_{-0.0012}$    & $9.1^{+0.3}_{-0.3}$       & $86.8^{+0.4}_{-0.4}$        &                           &                        \\
WASP-120 b                 & $0.07462^{+0.00054}_{-0.00052}$ & $6.62^{+0.21}_{-0.20}$    & $84.16^{+0.39}_{-0.38}$     & $0.21^{\rm a)}$           & $0.30^{\rm a)}$        \\
\multicolumn{1}{r}{Turner \etal (2016)} & $0.0809^{+0.0010}_{-0.0010}$    & $5.90^{+0.33}_{-0.33}$    & $82.54^{+0.78}_{-0.78}$     &                           &                        \\
\multicolumn{1}{r}{Alexoudi (2022)} & $0.0751^{+0.0005}_{-0.0005}$    & $6.80^{+0.20}_{-0.20}$    & $84.54^{+0.35}_{-0.35}$     &                           &                        \\
WASP-121 b                 & $0.11808^{+0.00057}_{-0.00047}$ & $3.776^{+0.034}_{-0.044}$ & $87.6^{+1.6}_{-1.2}$        & $0.24^{+0.04}_{-0.04}$    & $0.23^{+0.09}_{-0.08}$ \\
\multicolumn{1}{r}{Delrez \etal (2016)} & $0.12454^{+0.00047}_{-0.00048}$ & $3.754^{+0.023}_{-0.028}$ & $87.6^{+0.6}_{-0.6}$        &                           &                        \\
\multicolumn{1}{r}{Bourrier \etal (2020)} & $0.12355^{+0.00033}_{-0.00029}$ & $3.822^{+0.008}_{-0.008}$ & $89.10^{+0.58}_{-0.62}$     & $0.268^{+0.039}_{-0.039}$ & $0.14^{+0.08}_{-0.08}$ \\
\multicolumn{1}{r}{Wong \etal (2020)} & $0.12409^{+0.00045}_{-0.00043}$ & $3.815^{+0.018}_{-0.032}$ & $88.8^{+0.9}_{-1.2}$        & $0.251^{+0.045}_{-0.057}$ & $0.14^{+0.11}_{-0.08}$ \\
\multicolumn{1}{r}{Yang \etal (2022)} & $0.12340^{+0.00040}_{-0.00040}$ & $3.81^{+0.03}_{-0.03}$    & $89.9^{+1.6}_{-1.6}$        & $0.24^{+0.03}_{-0.03}$    & $0.21^{+0.04}_{-0.04}$ \\
\multicolumn{1}{r}{Daylan \etal (2021)} & $0.12488^{+0.00072}_{-0.00072}$ & $3.674^{+0.035}_{-0.035}$ & $85.26^{+0.64}_{-0.56}$     & $0.29^{+0.06}_{-0.06}$    & $0.06^{+0.11}_{-0.11}$ \\
WASP-122 b                 & $0.10999^{+0.00044}_{-0.00044}$ & $4.357^{+0.033}_{-0.037}$ & $78.88^{+0.13}_{-0.14}$     & $0.32^{\rm a)}$           & $0.27^{\rm a)}$        \\
\multicolumn{1}{r}{Turner \etal (2016)} & $0.1177^{+0.0012}_{-0.0012}$    & $4.248^{+0.072}_{-0.072}$ & $78.3^{+0.3}_{-0.3}$        &                           &                        \\
\multicolumn{1}{r}{Rodriguez \etal (2016)} & $0.1143^{+0.0029}_{-0.0026}$    & $4.64^{+0.25}_{-0.22}$    & $79.67^{+0.80}_{-0.77}$     &                           &                        \\
\multicolumn{1}{r}{Wong \etal (2020)} & $0.1102^{+0.0014}_{-0.0012}$    & $4.39^{+0.10}_{-0.09}$    & $78.91^{+0.47}_{-0.41}$     & $0.17^{+0.21}_{-0.12}$    & $0.21^{+0.23}_{-0.14}$ \\
WASP-140 b                 & $0.1363^{+0.0012}_{-0.0010}$    & $8.331^{+0.073}_{-0.075}$ & $84.07^{+0.08}_{-0.09}$     & $0.38^{\rm a)}$           & $0.24^{\rm a)}$        \\
\multicolumn{1}{r}{Hellier \etal (2017)} & $0.1432^{+0.0070}_{-0.0017}$    & $7.97^{+0.60}_{-0.26}$    & $83.3^{+0.5}_{-0.8}$        &                           &                        \\
\multicolumn{1}{r}{Alexoudi (2022)} & $0.1464^{+0.0010}_{-0.0010}$    & $8.58^{+0.06}_{-0.06}$    & $84.30^{+0.06}_{-0.06}$     &                           &                        \\
XO-6 b                     & $0.10747^{+0.00037}_{-0.00037}$ & $8.09^{+0.11}_{-0.09}$    & $84.85^{+0.13}_{-0.12}$     & $0.19^{\rm a)}$           & $0.32^{\rm a)}$        \\
\multicolumn{1}{r}{Crouzet \etal (2017)} & $0.110^{+0.006}_{-0.006}$       & $9.08^{+0.17}_{-0.17}$    & $86.0^{+0.2}_{-0.2}$        &                           &                        \\
\multicolumn{1}{r}{Ridden-Harper \etal (2020)} & $0.11494^{+0.00029}_{-0.00029}$ & $8.383^{+0.074}_{-0.074}$ & $85.24^{+0.09}_{-0.09}$     & $0.22^{\rm b)}$           & $0.32^{\rm b)}$        \\
XO-7 b                     & $0.09337^{+0.00049}_{-0.00048}$ & $6.19^{+0.10}_{-0.10}$    & $83.03^{+0.21}_{-0.21}$     & $0.25^{\rm a)}$           & $0.31^{\rm a)}$        \\
\multicolumn{1}{r}{Crouzet \etal (2020)} & $0.09532^{+0.00093}_{-0.00093}$ & $6.43^{+0.14}_{-0.14}$    & $83.45^{+0.29}_{-0.29}$     &                           &                        \\
\hline
\multicolumn{6}{l}{$R_{\rm{p}}/R_{\star}$, $a/R_{\star}$, and $i_{\rm{orb}}$ are the ratio of planet to star radii, semi-major axis scaled in star radii, and }  \\
\multicolumn{6}{l}{orbital inclination, respectively. $u_1$ and $u_2$ are the linear and quadratic LD coefficients. The values }  \\
\multicolumn{6}{l}{reported in this study are given in the lines with the planet names.}  \\
\multicolumn{6}{l}{$^{\rm a)}$ Value interpolated from Claret \& Bloemen (2011), varied under the Gaussian penalty.}  \\
\multicolumn{6}{l}{$^{\rm b)}$ Theoretical prediction fixed at a value from Claret (2017).}  \\
}

For the homogeneity of our transit timing studies, the fitting procedure was repeated for data publicly available in the literature. The new and re-determined mid-transit times are collected in Table~5.

\MakeTable{ l c c c c l}{12.5cm}{Transit mid-points for the studied planets.}
{\hline
Planet      & $E$ & $T_{\rm mid}$ (${\rm BJD_{TDB}}$) & $+\sigma$ (d) & $-\sigma$ (d) & Data source\\
\hline
HD 2685 b   & $0$ & $2458325.783762$ & $0.000370$ & $0.000391$ & this paper\\
HD 2685 b   & $1$ & $2458329.910256$ & $0.000411$ & $0.000410$ & this paper\\
HD 2685 b   & $2$ & $2458334.037635$ & $0.000381$ & $0.000402$ & this paper\\
\hline
\multicolumn{6}{l}{This table is available in its entirety in a machine-readable form at CDS. }  \\
\multicolumn{6}{l}{A portion is shown here for guidance regarding its form and content.}  \\
}

\subsection{Transit timing}

The mid-transit times were used to refine linear ephemerides for the individual planets in the form:
\begin{equation}
     T_{\rm mid }(E) = T_0 + P_{\rm orb} \cdot E \, , \;
\end{equation}
where $E$ is a transit number counted from a reference epoch $T_0$, taken from a discovery paper. The MCMC algorithm was employed to derive the best-fitting parameters and their $1\sigma$ uncertainties from posterior probability distributions produced by 100 chains, each of them was $10^4$ steps long with the first 1000 trials discarded. The results are given in Table~6, and the timing residuals against the refined ephemerides are plotted in Figs.~2, 3, and 4.

\MakeTable{ l c c }{12.5cm}{Transit ephemeris elements for the investigated planets.}
{\hline
Planet      & $T_0$ (${\rm BJD_{TDB}}$) & $P_{\rm orb}$ (d)\\
\hline
HD 2685 b   & $2458325.78366 \pm 0.00015$ & $4.1269052 \pm 0.0000011$ \\
Qatar 10 b  & $2458247.90724 \pm 0.00029$ & $1.64532648 \pm 0.00000055$ \\
WASP-4 b    & $2454368.59335 \pm 0.00003$ & $1.33823137 \pm 0.00000002$ \\
WASP-48 b   & $2455364.55109 \pm 0.00028$ & $2.14363708 \pm 0.00000023$ \\
WASP-58 b   & $2455183.9342 \pm 0.0010$   & $5.0172130 \pm 0.0000012$ \\
WASP-91 b   & $2456297.71905 \pm 0.00019$ & $2.79857960 \pm 0.00000024$ \\
WASP-120 b  & $2456779.43545 \pm 0.00067$ & $3.6112673 \pm 0.0000012$ \\
WASP-121 b  & $2456635.70841 \pm 0.00014$ & $1.27492471 \pm 0.00000008$ \\
WASP-122 b  & $2456665.22476 \pm 0.00029$ & $1.71005327 \pm 0.00000022$ \\
WASP-140 b  & $2456912.35143 \pm 0.00028$ & $2.23598468 \pm 0.00000036$ \\
XO-6 b      & $2456652.71351 \pm 0.00050$ & $3.7649930 \pm 0.0000008$ \\
XO-7 b      & $2457917.47508 \pm 0.00041$ & $2.8641362 \pm 0.0000010$ \\
\hline
}

\begin{figure}[thb]
\begin{center}
\includegraphics[width=1.0\textwidth]{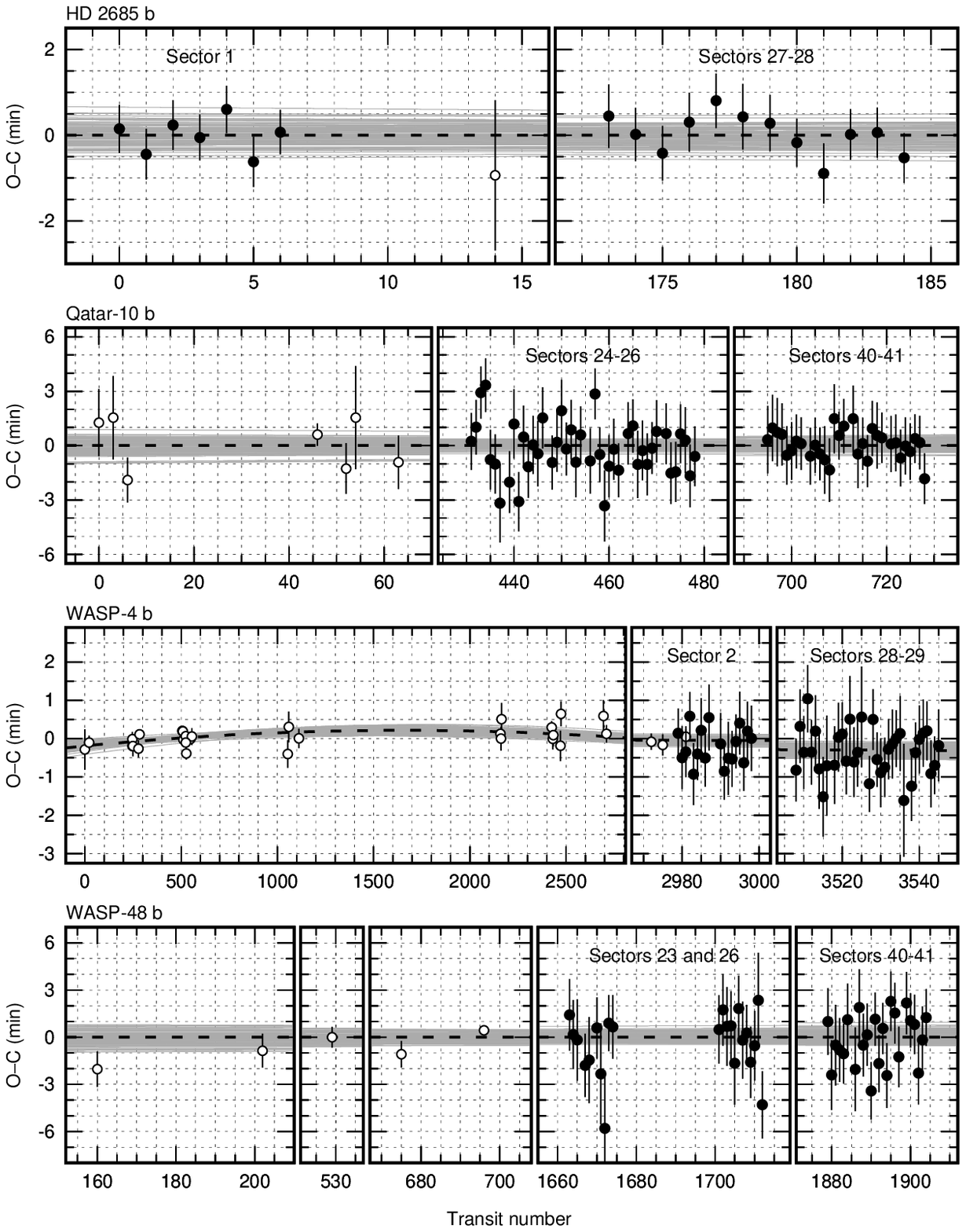}
\end{center}
\FigCap{Transit-timing residuals against the refined linear ephemerides for HD 2685~b, Qatar-10~b, WASP-4~b, and WASP-48~b. The values from TESS photometry are marked with dots, and the redetermined literature values are plotted with open circles. Dashed lines mark zero value. For WASP-4~b, the dashed line shows the quadratic trend in transit times. The ephemeris uncertainties are illustrated by grey lines plotted for 100 sets of parameters drawn from the Markov chains.}
\end{figure}

\begin{figure}[thb]
\begin{center}
\includegraphics[width=1.0\textwidth]{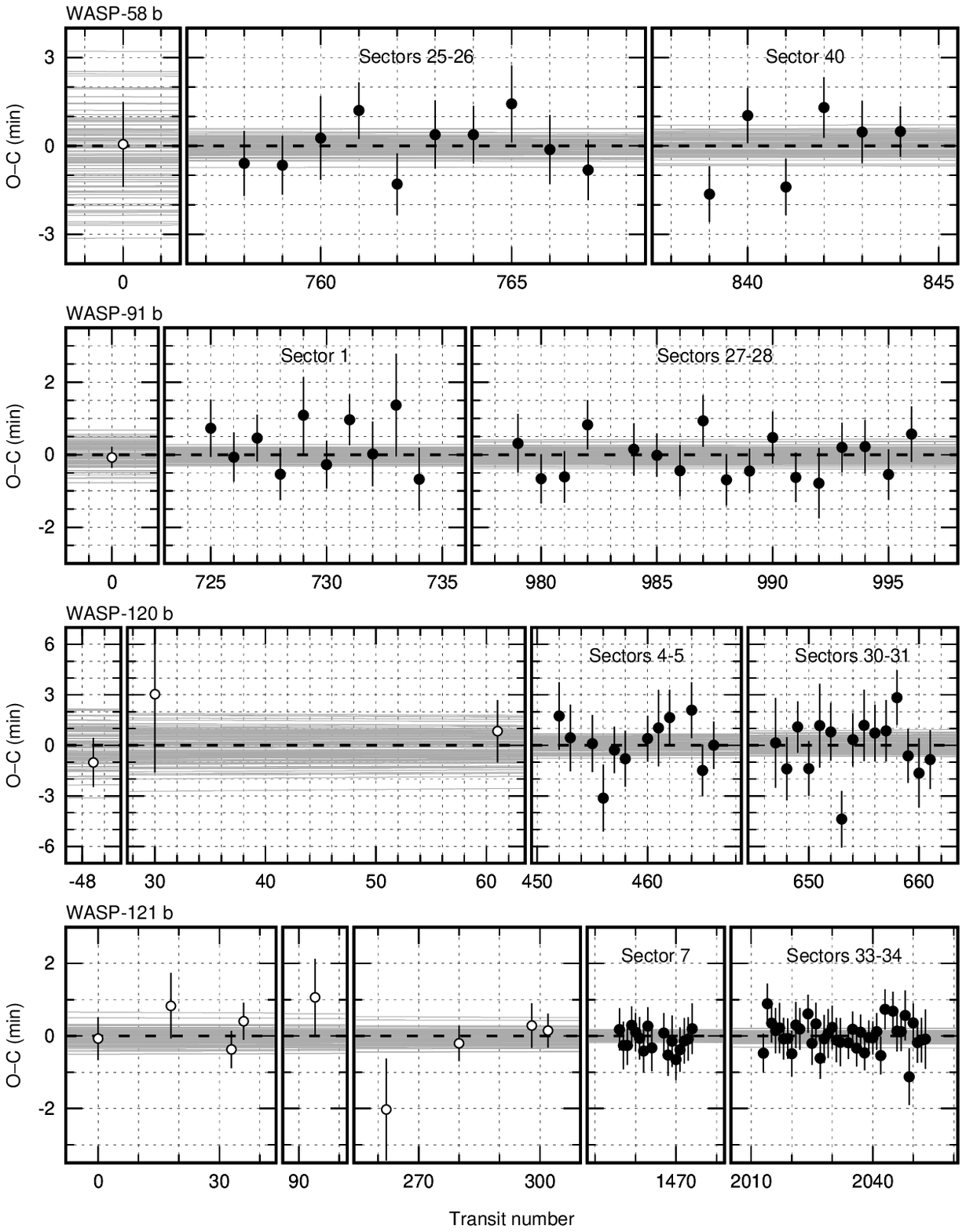}
\end{center}
\FigCap{The same as Fig.~2 but for WASP-58~b, WASP-91~b, WASP-120~b, and WASP-121~b.}
\end{figure}

\begin{figure}[thb]
\begin{center}
\includegraphics[width=1.0\textwidth]{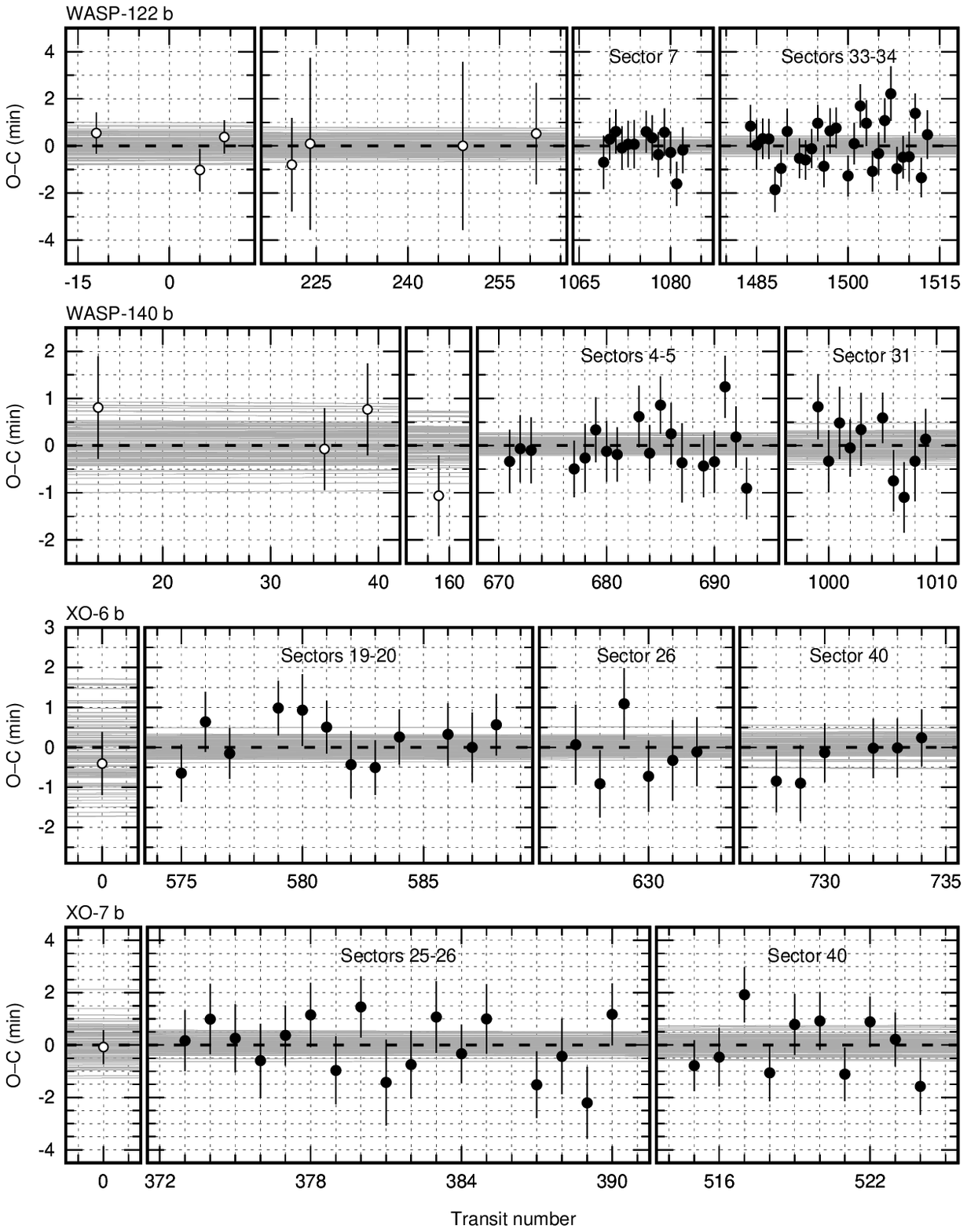}
\end{center}
\FigCap{The same as Fig.~2 but for WASP-122~b, WASP-140~b, XO-6~b, and XO-7~b.}
\end{figure}

To probe possible long-term trends which could be caused by a monotonic or periodic change of $P_{\rm orb}$, trial quadratic ephemerides in the form:
\begin{equation}
 T_{\rm{mid}}= T_0 + P_{\rm{orb}} \cdot E + \frac{1}{2} \frac{{\rm d} P_{\rm{orb}}}{{\rm d} E} \cdot E^2 \, , \;
\end{equation}
where ${{\rm d} P_{\rm{orb}}}/{{\rm d} E}$ is the change in the orbital period between succeeding transits, were tested. The Bayesian information criterion (BIC) disfavours the quadratic ephemerides for all planets of our sample but WASP-4~b. For that system, we derived ${{\rm d} P_{\rm{orb}}}/{{\rm d} E} = (-2.1 \pm 0.6) \cdot 10^{-10}$ days per orbit with $\Delta \rm{BIC} = {\rm{BIC}}_{\rm{linear}} - {\rm{BIC}}_{\rm{quadratic}} \approx 6.7$. In Fig.~2, the quadratic trend is plotted after subtraction of the linear ephemeris. 

The timing data were searched for periodic signals after subtracting the quadratic and linear ephemerides for WASP-4~b and the remaining planets, respectively. The analysis of variance algorithm (AoV, Schwarzenberg-Czerny 1996) was used to calculate periodograms for trial periods in a range of $2-10^4$ epochs. As displayed in Fig.~5, no statistically significant signal was detected for any planet. The bootstrap method, based on $10^5$ trials, was employed to determine the levels of the false alarm probability (FAP) empirically.

\begin{figure}[thb]
\begin{center}
\includegraphics[width=1.0\textwidth]{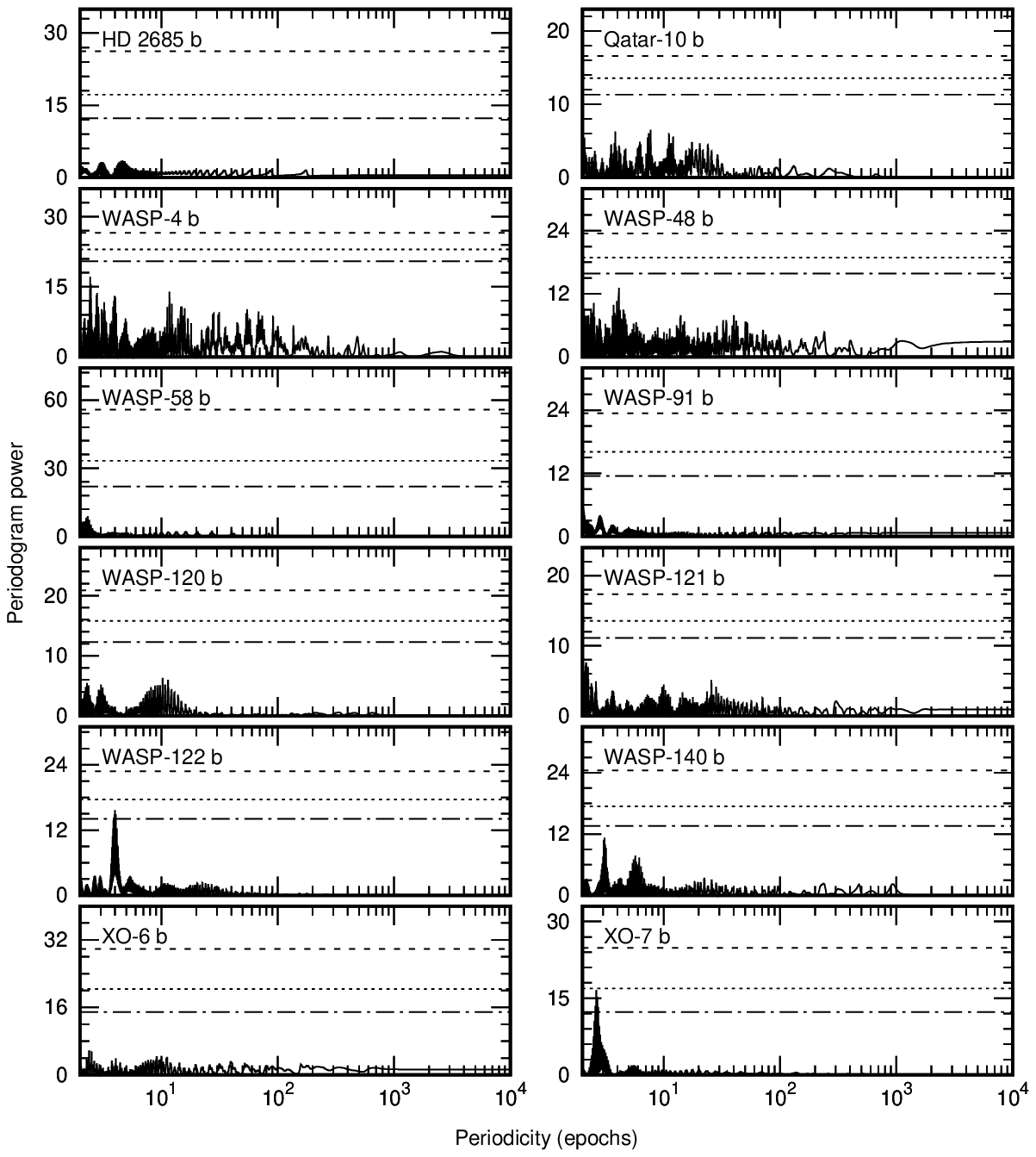}
\end{center}
\FigCap{AoV periodograms for the timing residuals against the redetermined ephemerides. The dashed and dotted horizontal lines show the empirical FAP levels of 5\%, 1\%, and 0.1\% (from the bottom up).}
\end{figure}

\subsection{Search for additional transiting planets}

As in Maciejewski (2020), the AoV method designed to detect transit-like signals (AoVtr, Schwarzenberg-Czerny \& Beaulieu 2006) was fed with the photometric time series. The transits and occultations of the known planets were masked out. Box-like periodic signals were searched in a range of $0.5-100$ days with a resolution in the frequency domain equal to $5 \times 10^{-5}$ day$^{-1}$. The periodograms with the FAP levels, determined with the bootstrap method on $10^4$ resampled datasets, are plotted in Fig.~6. Our analysis revealed no statistically significant signals.

Artificial transits were injected into the original light curves to determine transit detection thresholds. The transit depths at which FAP dropped to 0.1\% were recorded for trial orbital periods between $0.5-100$ days. The results are over-plotted in Fig.~6 for the individual systems. Median transit depths for orbits with periods below 10 days are collected in Table~7. They were translated into the upper radii of hypothetical planets that remain below the detection thresholds of the observations.

\begin{figure}[thb]
\begin{center}
\includegraphics[width=0.97\textwidth]{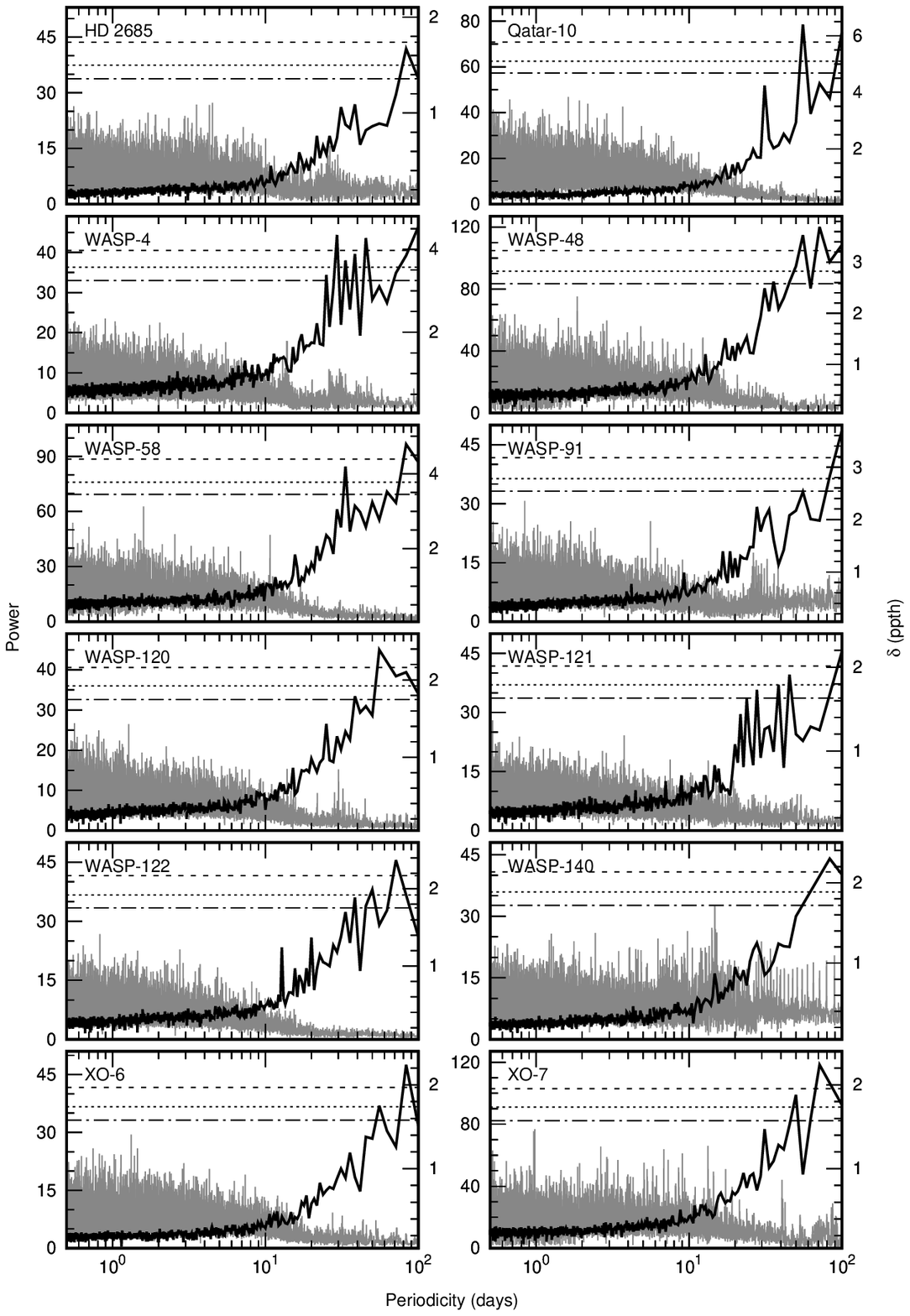}
\end{center}
\FigCap{AoVtr periodograms (on left axes) and upper constraints on depths of transits that avoid detection (right axes) in the examined systems. The power spectra are drawn with grey lines. The dashed and dotted horizontal lines mark the empirical FAP levels of 5\%, 1\%, and 0.1\% (from the bottom up). The empirical constraints on the transit depths are plotted with black bold lines.}
\end{figure}

\MakeTable{ l c c}{12.5cm}{Summary of transit detection sensitivity.}
{\hline
System   & $\delta_{\rm lim}$ (ppth) & $R_{\rm lim}$ $(R_{\oplus})$\\
\hline
HD2685   & 0.18 & $2.29 \pm 0.07$ \\
Qatar-10 & 0.40 & $2.74 \pm 0.06$ \\
WASP-4   & 0.62 & $2.55^{+0.11}_{-0.08}$ \\
WASP-48  & 0.42 & $3.9 \pm 0.2$ \\
WASP-58  & 0.57 & $3.05 \pm 0.34$ \\
WASP-91  & 0.39 & $1.85 \pm 0.07$ \\
WASP-120 & 0.29 & $3.5 \pm 0.2$ \\
WASP-121 & 0.29 & $2.71 \pm 0.06$ \\
WASP-122 & 0.29 & $2.83 \pm 0.06$ \\
WASP-140 & 0.26 & $1.69 \pm 0.05$ \\
XO-6     & 0.20 & $3.0 \pm 0.3$ \\
XO-7     & 0.25 & $2.56 \pm 0.04$ \\
\hline
\multicolumn{3}{l}{$\delta_{\rm lim}$ is the median of the limiting transit depths for the orbital periods below 10 days.}  \\
\multicolumn{3}{l}{The values of $\delta$ were converted into planetary radii $R_{\rm lim}$ using stellar radii from Table 2.}  \\
}


\section{Discussion}

The hot Jupiters presented in this sample were found to be devoid of nearby planetary companions. This finding aligns with conclusions presented recently by Hord \etal (2021) for a sample of 184 hot Jupiters in the southern ecliptic hemisphere. The transit method allowed us to probe down to the sub-Neptunian regime of planetary radii. In the systems WASP-91 and WASP-140, a level of terrestrial-like planets ($<1.7$ $R_{\oplus}$, Buchhave \etal 2014) was reached. We notice, however, that hypothetical smaller planets, such as super-Earths WASP-47 e (1.8 $R_{\oplus}$, Becker \etal 2015) or Kepler-730 c (1.6 $R_{\oplus}$, Ca{\~n}as \etal 2019), are likely missed. They could be detected with larger space-borne instruments such as CHaracterizing ExOPlanets Satellite (CHEOPS, Broeg \etal 2013). The lack of periodic transit timing variations demonstrates that the investigated planets have no non-transiting companions such as WASP-148~c (H\'ebrard \etal 2020) that would stay close to the orbital period commensurabilities.

As demonstrated by Baluev \etal (2019), the rate of the change in the orbital period of WASP-4~b and its significance depend on a set of timing data considered. The value of ${{\rm d} P_{\rm{orb}}}/{{\rm d} E}$ can be transformed into $\dot{P}_{\rm{orb}}$ using the relation:
\begin{equation}
\dot{P}_{\rm{orb}} = \frac{1}{P_{\rm{orb}}} \frac{{\rm d} P_{\rm{orb}}}{{\rm d} E} \, . \;
\end{equation} 
Bouma \etal (2019) reported $\dot{P}_{\rm{orb}} = -12.6 \pm 1.2$ ms~yr$^{-1}$, and then refined to $-8.6 \pm 1.3$ ms~yr$^{-1}$ (Bouma \etal 2020). The latter result is in line with $-9.2 \pm 1.1$ ms~yr$^{-1}$ obtained by Southworth \etal (2019). Thanks to additional TESS observations, hence a longer time-baseline, Turner \etal (2021) obtained $-7.33 \pm 0.71$ ms~yr$^{-1}$. Baluev \etal (2020) performed a homogenous re-analysis of the transit light curves, including those from Hoyer \etal (2013) and Huitson \etal (2017) that are not publicly available, and found $\dot{P}_{\rm{orb}} = -5.4 \pm 1.5$ ms~yr$^{-1}$, \ie a substantially weaker trend with just 3.6$\sigma$ significance. Our timing observations cover the same time as Turner \etal (2021) data, but do not contain the data from Hoyer \etal (2013) and Huitson \etal (2017). We obtained $\dot{P}_{\rm{orb}} = -4.8 \pm 1.4$ ms~yr$^{-1}$. As discussed in Turner \etal (2021), the nature of the observed departure from a linear ephemeris for transits of WASP-4~b remains unclear. The orbital decay and apsidal precession scenarios are considered and future observations are required to shed more light on this matter.

Our redetermined values of the orbital inclination, semi-major axis scaled in star radii, and LD coefficients (for HD 2685, WASP-4, and WASP-121) agree with the literature determinations within $1-2\,\sigma$. The only exception is XO-6~b: its orbital inclination was found to be smaller by 4.8$\sigma$ and scaled semi-major axis smaller by 4.9$\sigma$ when compared to the values reported by Crouzet \etal (2017). This finding is confirmed by Ridden-Harper \etal (2020) results, also based on TESS data. For six planets of our sample, the relative radii were found to be significantly underestimated. Visual inspection of sky fields around those targets revealed that relatively bright neighbour stars contaminated their fluxes. However, taking that effect into account is out of the scope of this study. For five systems: HD~2685, Qatar-10, WASP-58, WASP-91, and XO-7, we refined their parameters for the first time.

\section{Conclusions}

With the 12 planetary systems analysed in this paper, we expanded the sample of systems studied in our project to 20 in total. Hord \etal (2021) predict that the occurrence rate for hot Jupiters in compact systems is as low as $4.2^{+9.1}_{-4.2}$\% if only companions with radii $>2$ $R_{\oplus}$ are considered. The negative results of our survey are in line with these statistics and support the high-eccentricity migration as a pathway of those giant planets to the tight orbits observed today. We notice that the TESS photometric data allowed us to explore down to the sub-Neptunian sizes, leaving, however, the regimes of terrestrial planets and mini-Neptunes untouched.


\Acknow{We are grateful to Dr.~Laetitia Delrez, Dr.~Coel Hellier, and Dr.~Zlatan Tsvetanov for sharing the follow-up light curves with us. We are also grateful to the anonymous referee for the careful reading of the manuscript. This paper includes data collected with the TESS mission, obtained from the MAST data archive at the Space Telescope Science Institute (STScI). Funding for the TESS mission is provided by the NASA Explorer Program. STScI is operated by the Association of Universities for Research in Astronomy, Inc., under NASA contract NAS 5-26555. This research made use of Lightkurve, a Python package for Kepler and TESS data analysis (Lightkurve Collaboration, 2018). This research has made use of the SIMBAD database and the VizieR catalogue access tool, operated at CDS, Strasbourg, France, and NASA's Astrophysics Data System Bibliographic Services.}


\end{document}